\documentclass{ws-jmrr}
\usepackage[sort,compress,super]{cite}
\usepackage{amsmath,amssymb,amsfonts}
\usepackage{algorithmic}
\usepackage{graphicx}
\usepackage{textcomp}
\usepackage{wrapfig}
\usepackage{enumitem}
\usepackage{caption}
\usepackage{subcaption}
\usepackage{epsfig}
\usepackage{multirow}
\usepackage{derivative}
\usepackage{siunitx}
\usepackage{xcolor, soul}
\usepackage{outlines}
\usepackage{bigints}
\usepackage{hyperref}

\hypersetup{nolinks=true}

\newcommand{\B}[1]{\ensuremath{\mathbb{#1}}}

\newcommand{\reals}{\B{R}}

\newcommand{\abs}[1]{\ensuremath{\left\lvert #1 \right\rvert}}
\newcommand{\norm}[1]{\ensuremath{\left\lVert #1 \right\rVert}}
\newcommand{\lnorm}[2][2]{
    \ifthenelse{ \equal{#1}{1} }
    {\abs{#2}}
    {\norm{#2}_{#1}}
}
\newcommand{\degrees}{^\circ}
\newcommand{\etal}{\textit{et. al}}
\renewcommand{\unit}[1]{~\si{#1}}
\newcommand{\selfnote}[2][]{
    \ifthenelse{ \equal{#1}{done} }
        { \emph{\st{\MakeUppercase{#2}}} }
        { \hl{\emph{\MakeUppercase{#2}}} }
} 
\newcolumntype{L}{>{$}l<{$}} 
\newcolumntype{C}{>{$}c<{$}} 
\newcommand{\winit}[1][]{
    \ensuremath{
        \ifthenelse{ \equal{#1}{} }
        { \boldsymbol{\omega}_{init} }
        { \boldsymbol{\omega}_{init, #1} }
    }
}
\newcommand{\kappac}[1][]{
    \ensuremath{
        \ifthenelse{ \equal{#1}{} }
        { \kappa_{c} }
        { \kappa_{c, #1} }
    }
}
\newcommand{\Eqref}[1]{Eq. \eqref{#1}}
\newcommand{\exvivo}{\textit{ex-vivo}~}
\begin{document}


\newcommand{\papertitle}{Optical Fiber-Based Needle Shape Sensing in Real
Tissue: Single Core vs. Multicore Approaches}

\markboth{D. A. Lezcano, \etal }{\papertitle}

\title{\papertitle}
\renewcommand*{\thefootnote}{\fnsymbol{footnote}}

\author{
    Dimitri A. Lezcano$^a$, 
    Yernar Zhetpissov$^{a}\footnote{Yernar Zhetpissov was affiliated with Johns Hopkins University at the time of this work and is now currently affiliated Worcester Polytecnic Institute's Robotics Engineering Department. }$, 
    Alexandra Cheng$^b$, 
    Jin Seob Kim$^a$, 
    Iulian I. Iordachita$^a$
}

\address{$^a$Mechanical Engineering, 
Johns Hopkins University, 
3400 North Charles St.,
Baltimore, Maryland 21218, United States\\
E-mail: \{dlezcan1, yzhetpi1, jkim115, iordachita\}@jhu.edu}


\address{$^b$Biomedical Engineering, 
Johns Hopkins University, 
3400 North Charles St.,
Baltimore, Maryland 21218, United States\\
E-mail: hcheng33@jhu.edu}

\maketitle

\begin{abstract}
Flexible needle insertion procedures are common for minimally-invasive surgeries for diagnosing and treating prostate cancer.
Bevel-tip needles provide physicians the capability to steer the needle during long insertions to avoid vital anatomical structures in the patient and reduce post-operative patient discomfort.
To provide needle placement feedback to the physician, sensors are embedded into needles for determining the real-time 3D shape of the needle during operation without needing to visualize the needle intra-operatively.
Through expansive research in fiber optics, a plethora of bio-compatible, MRI-compatible, optical shape-sensors have been developed to provide real-time shape feedback, such as single-core and multicore fiber Bragg gratings. 
In this paper, we directly compare single-core fiber-based and multicore fiber-based needle shape-sensing through identically constructed, four-active area sensorized bevel-tip needles inserted into phantom and \exvivo tissue on the same experimental platform.
In this work, we found that for shape-sensing in phantom tissue, the two needles performed identically with a $p$-value of $0.164 > 0.05$, but in \exvivo real tissue, the single-core fiber sensorized needle significantly outperformed the multicore fiber configuration with a $p$-value of $0.0005 < 0.05$.
This paper also presents the experimental platform and method for directly comparing these optical shape sensors for the needle shape-sensing task, as well as provides direction, insight and required considerations for future work in constructively optimizing sensorized needles. 
\end{abstract}


\begin{multicols}{2}
\flushcolumns

\section{Introduction}\label{sec:intro}
\par Bevel-tip needle insertion procedures are pervasive surgical techniques for minimally-invasive surgeries, including but not limited to biopsy, brachytherapy, and prostate cryoablation \cite{Okazawa2005, Matlaga2003, Pernar2018}. 
Bevel-tip needles enable needle steering, leveraging the asymmetric force distribution applied to the tip of the needle during needle insertion.
Due to randomness observed in surgeries such as tissue obstruction, movement of patient, or practitioner error a needle may deviate from its intended trajectory.
From these challenges, corrections to the needle's trajectory are typically performed from repetitive reinsertions into the patient, angulating the needle to better reach the needle's intended target.  
These reinsertions cause unnecessary tissue damage to the patient, damaging nearby sensitive anatomical structures and resulting in post-operative patient discomfort \cite{Loeb2011}. 
Thus, guidance solutions are imperative to mitigate needle insertion error, providing real-time feedback to the needle's location in the patient, minimizing risk and improving patient outcomes \cite{Glozman2007}. 
Conventional methods for surgical guidance include real-time imaging modalities such as ultrasound imaging \cite{Adebar2014, Shahriari2016, Chen2021, Mathews2022, Rabiei2022} and MRI \cite{Su2011, Li2020a, AlMaatoq2023} for tracking the needle’s trajectory during the needle insertion. 
CT is another modality for imaging and tracking the needle's trajectory, however is not usually real-time and requires high doses of radiation to be delivered to the patient \cite{Medan2019, Reisenauer2022, Wei2023}.
An alternative approach utilizes needles embedded with fiber-Bragg grating (FBG) sensors \cite{Roesthuis2015, Shahriari2016, Zhang2019, Li2006, VandeBerg2015, Donder2021, Zhang2023}. 
FBG fibers are optical sensors capable of detecting locally induced strain derived from Bragg's law using peak backscattered optical wavelength at the sensing locations along the fiber, denoted as \emph{active areas (AAs)}. 
Using the local curvature estimates from the FBGs along the needle, the needle's shape can be accurately estimated without direct observation of the needle using an imaging modality. 
Furthermore, FBG sensors are MRI-compatible, allowing for shape-sensing to be used in conjunction with the aforementioned imaging modalities \cite{Park2010, Seifabadi2013, Shahriari2016, Ourak2021}.

\par Currently, standardized FBG-sensorized bevel-tip needles are not readily found on the market, thus requiring for these devices to be built by individuals according to their own requirements. 
Without a standardized needle, optimizing the needle's hardware design becomes an important topic of research for the development of any FBG shape-sensing needles. 
Previous works into the optimization of sensorized needle construction have ranged from sensor placement to experiments leveraging fiber imperfections for improving shape-sensing accuracy \cite{Zhou2018, Al-Ahmad2020, Kim2017, Zhang2019, Lehocky2014, Idrisov2021, Deaton2023}. 
These efforts typically propose a single needle hardware design and discuss the optimization methods of such design in comparison with other works.

\par Through the advancement of optical sensor fabrication technology, several novel variants of FBG sensor have emerged, providing researchers different options for choosing shape sensors. 
In this work, we fabricated two needles with identical form factors and similar sensor structure but used two different variants of FBG sensors: one using single-core FBGs (SCFs) and the other with a multicore FBG (MCF) \cite{Lai2020, Lu2021}. 
Other ongoing research in optical shape-sensors include using distributed FBGs and advanced signal processing methods in SCFs and MCFs for 3D shape-sensing methods, improving upon discrete FBG placement, however typically require specialized interrogators for processing distributed FBG signals \cite{Zhao2016a, Beisenova2019, Amantayeva2021, Francoeur2023}. 
Given all of the research to develop novel shape sensors, there exists a gap in current state-of-art evaluating these directly to each other for needle shape-sensing.
Considerations needed for sensorized needles for shape estimation require to be cost-effective, real-time, bio-compatible, and reliable.
Therefore, the need for direct comparisons of these sensing modalities is imperative to determine the optimal construction of shape-sensing needles.

\par In this work, we provide a baseline evaluation comparison of two identically configured needles, one embedded with SCFs and the other with an MCF, for the needle shape-sensing task through needle insertions into phantom and \exvivo tissue.
Furthermore, this work develops an evaluation platform for direct comparison of future sensing modalities for needle shape-sensing and provides direction for future research in constructive optimization for MCF-based needle shape-sensing.
For three-dimensional SCF shape-sensing, at least two channels non-$180\degrees$ increments of each other are required within one cross-section \cite{Roesthuis2014}.
We incorporated a third channel in our SCF needle design to enable temperature-invariant shape sensing. 
The two needles realize an identical channel orientation, with the three channels lying on the same circle and $120\degrees$ apart from one another, but at varying radial distances from the needle's central axis.
There exists a central core in the MCF needle lying along the needle's central axis, typically used for temperature compensation \cite{Paloschi2021}.
The novelty of this work includes a direct performance comparison of SCF-sensorized and MCF-sensorized needles in phantom and \exvivo tissue, a presentation of possible sources of errors for using SCFs and MCFs as needle shape sensors, and identification of future research directions in optimally constructing MCF-sensorized needles for shape-sensing.

\section{Needle Construction}\label{sec:needles}
\par We fabricated two MRI-compatible identical 18G (OD $\sim$ 1.3 mm) needles that are 200 mm in length (KIM18/20, ITP GmbH, Bochum, Germany), but fabricated one with three SCF sensors and the other with an MCF sensor. 
Each of the sensors had four FBG AAs, identically located at points along the needle, as shown in Fig. \ref{fig:needles_config_overall}.

\begin{figurehere}
    \begin{subfigure}[b]{\linewidth}
        \includegraphics[width=0.9\textwidth]{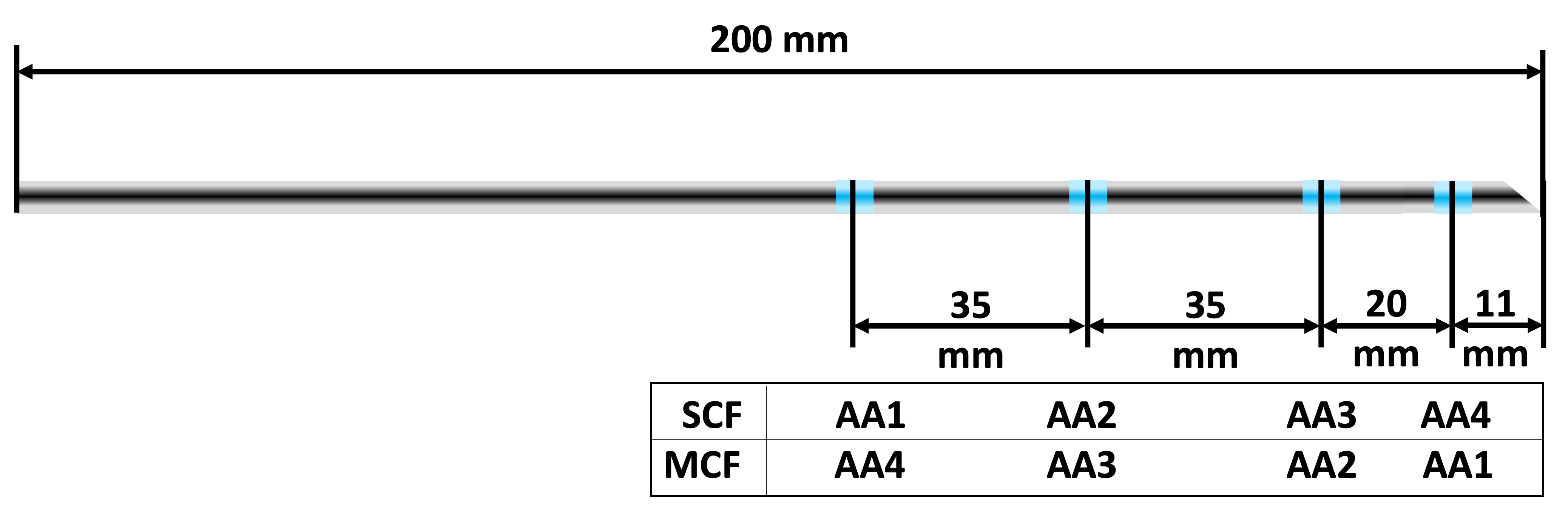}
        \caption{FBG placement}
        \label{fig:needles_config_overall}
    \end{subfigure}
    \begin{subfigure}[b]{0.45\linewidth}
        \centering
        \includegraphics[height=3.0cm, width=\textwidth]{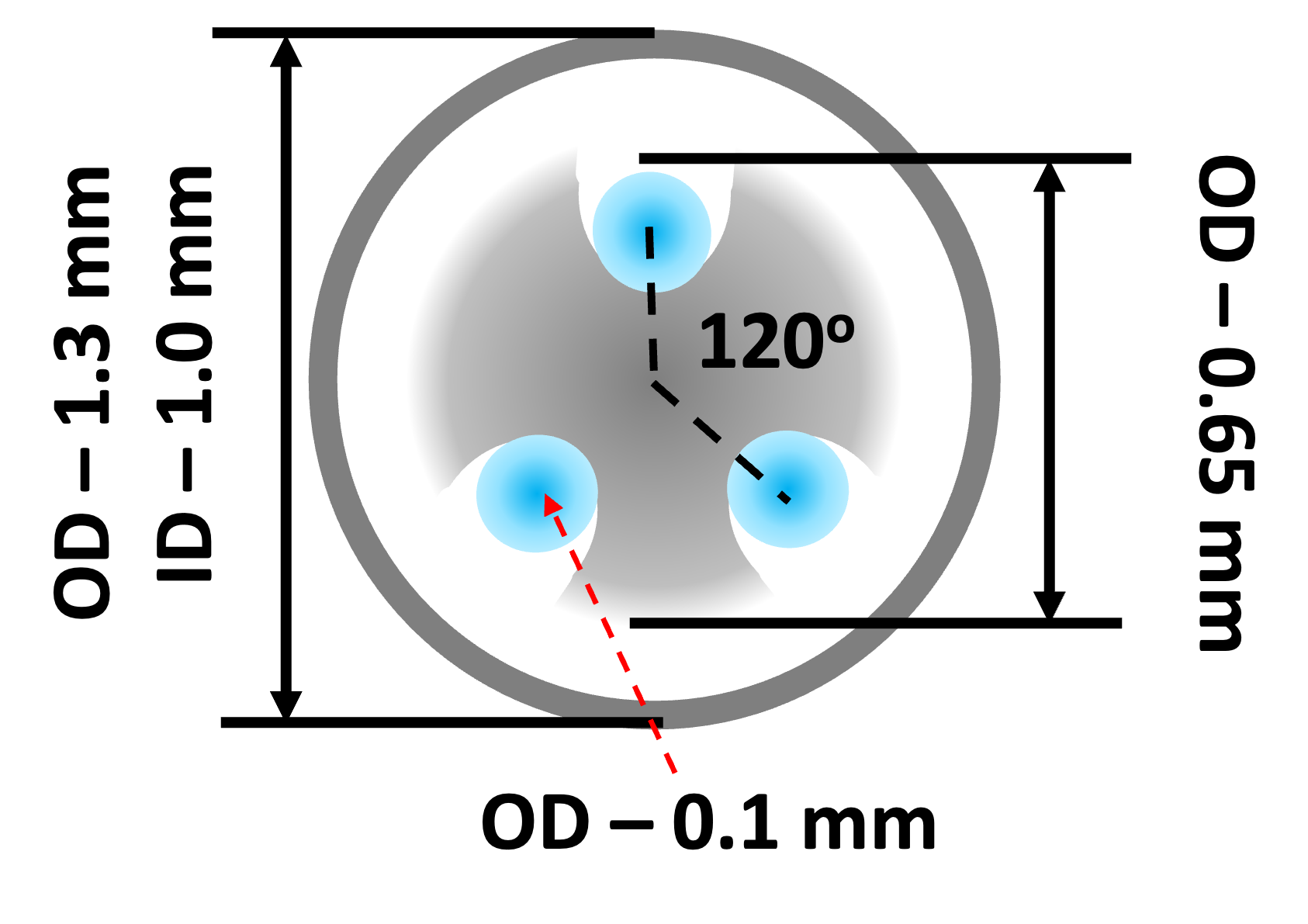}
        \caption{SCF}
        \label{fig:needles_config_scf}
    \end{subfigure}
    \begin{subfigure}[b]{0.225\textwidth}
        \includegraphics[height=3.0cm, width=\textwidth]{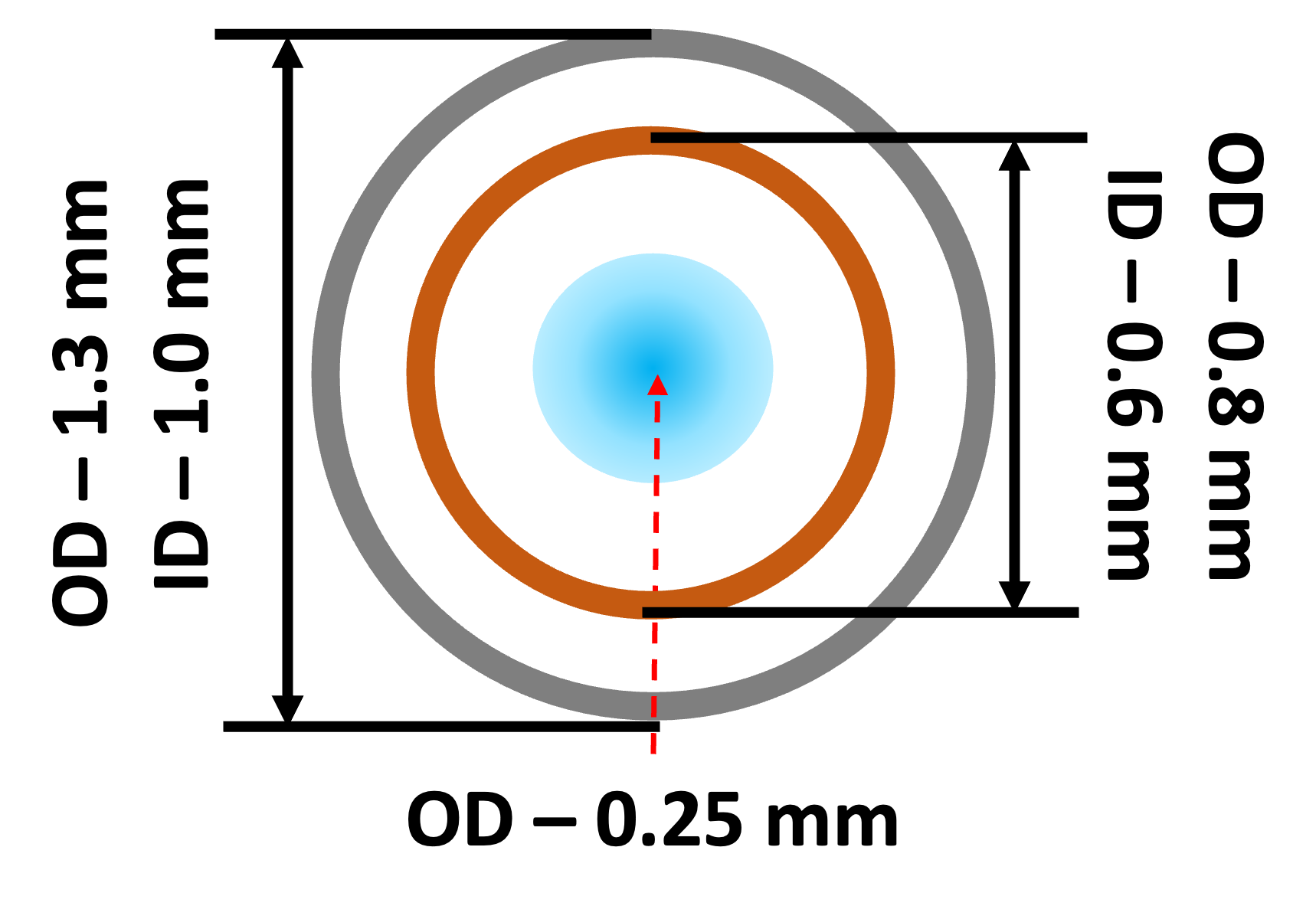}
        \caption{MCF}
        \label{fig:needles_config_mcf}
    \end{subfigure}
    \caption{The sensor configuration for the two 18G sensorized needles used in this work. 
    (\subref{fig:needles_config_overall}) The FBG placements and AA numbering along both the SCF and MCF needles.
    (\subref{fig:needles_config_scf}) and (\subref{fig:needles_config_mcf}) are the cross-sections of the SCF and MCF needles, respectively.
    The blue points mark the FBG active areas in (\subref{fig:needles_config_overall}) and the fiber optic cables containing FBG active areas in (\subref{fig:needles_config_scf}) and (\subref{fig:needles_config_mcf}).
    }
    \label{fig:needles_config}
\end{figurehere}

\subsection{Three-Channel Single-Core Fiber Needle}\label{sec:needles_scf}
\par For the SCF needle, there are three single-core fibers (80-micron cladding diameter, Technica Optical Components LLC, Atlanta, GA) embedded at $120\degrees$ increments from each other, illustrated in Fig. \ref{fig:needles_config_scf}.
The fibers were glued to a nitinol inner stylet's grooves with a bio-compatible adhesive (Loctite AA 3322, Henkel, Rocky Hill, CT). 
Glue was applied and cured in 3-4 mm increments along the needle for each of the three SCFs to ensure proper adhesion of the sensors to the needle.
The entire SCF needle construction process took longer than four hours to complete and was followed with an overnight cure.
Since the most bending that will naturally occur with the bevel-tip needle aligns in the direction of the bevel, we ensured that one of the SCFs embedded in the needle align directly in the direction of the bevel, to maximize strain experienced in one of these fibers and ensuring that the other two fibers experience strain.

\subsection{Multicore Fiber Needle}\label{sec:needles_mcf}
\begin{figurehere}
    \centering
    \includegraphics[width=0.5\linewidth]{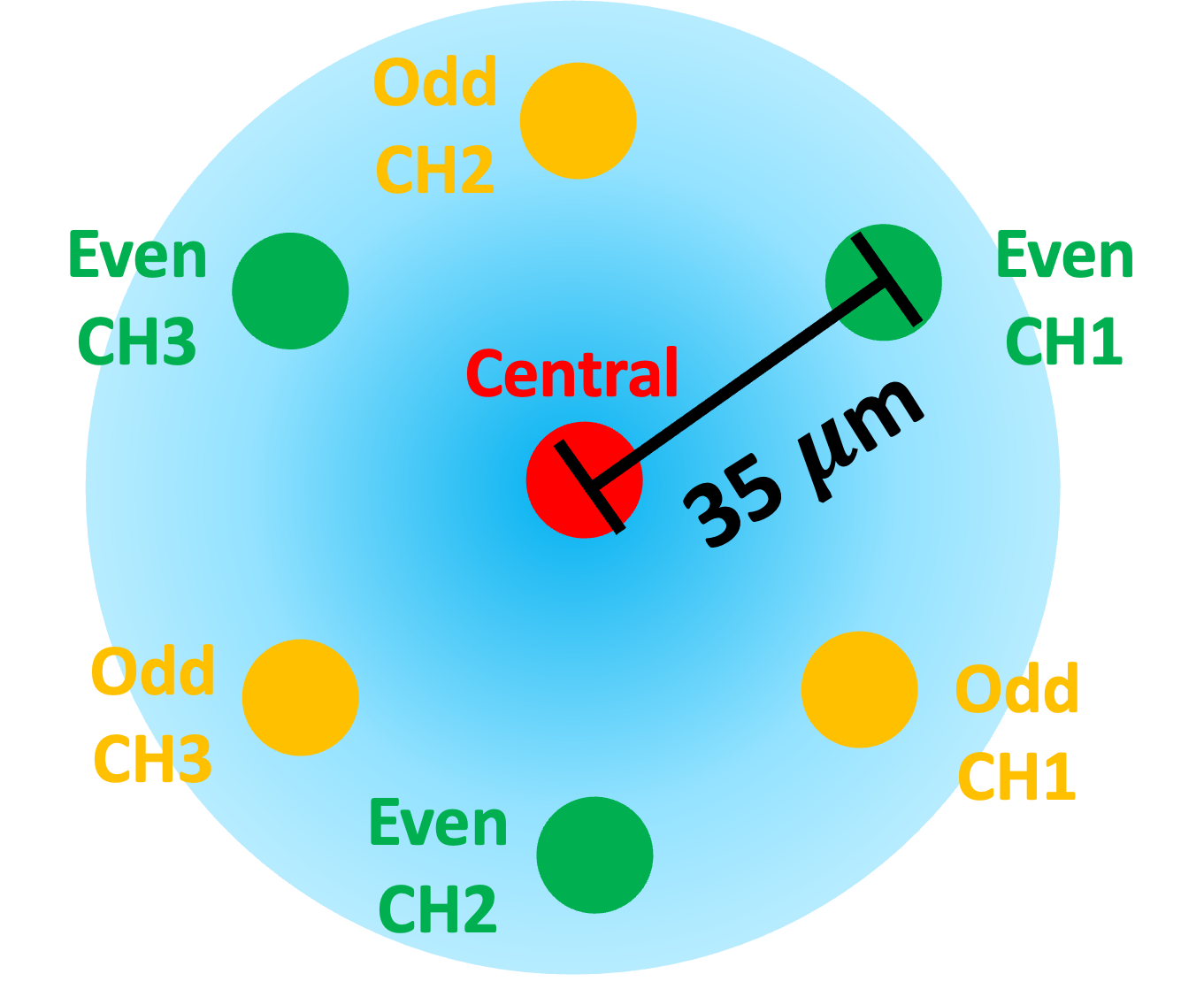}
    \caption{Cross-section of the multicore fiber containing seven cores, six outer cores and a central core. In this work, we use the even cores and the central core.}
    \label{fig:needles_mcf_channels}
\end{figurehere}
\par The MCF needle required only a single fiber-optic cable attached with a fanout. 
The MCF sensor (125-micron cladding diameter, Fujikura America, Sunnyvale, CA) had seven channels embedded into a single fiber-optic cable, one central core channel at the center of the cable and six outer channels in a hexagonal pattern around the central core, shown in Fig. \ref{fig:needles_mcf_channels}.
In this experiment, we used the even cores and the central core for shape-sensing where our decision was informed by [\citenum{Cheng2023}].
Different from the SCF needle, the MCF needle had an additional inner stylet to increase the strain transfer from the needle's outer stylet, illustrated in Fig. \ref{fig:needles_config_mcf}. 
The additional inner stylet was imperative since the single fiber-optic cable was placed close to the needle's central axis, which reduced the amount of strain induced for the same curvature, resulting in lower sensor sensitivity. 
The MCF sensor was mounted to the needle by gluing one end of the cable to the base of the needle, ensuring the sensor was placed coaxially with the needle. 
Similarly to the SCF needle, two channels were placed in the natural bending plane of the needle, aligned with the needle's bevel-tip for the same reasons listed in Sec. \ref{sec:needles_scf}.

\section{Models and Methods}\label{sec:methods}
\subsection{FBG Sensor Model}\label{sec:methods_fbg}
\par FBG sensors are capable of detecting curvature through strain measurements from shifts in the sensor's Bragg wavelength, $\lambda_B$. Due to the periodicity of the grating in the fiber, the unstrained Bragg wavelength, $\lambda_{B, 0}$, will shift from a change in strain, $\Delta\epsilon$, and temperature $\Delta T$ according to
\begin{equation}\label{eq:methods_fbg_bragg}
    \frac{\Delta \lambda_B}{\lambda_{B, 0}} = S_\epsilon \Delta \epsilon + S_T \Delta T,
\end{equation}
where $S_\epsilon$ and $S_T$ are the strain and temperature sensitivity coefficients of the grating, respectively.

\par Applying Euler-Bernoulli beam theory, we have that the strain measured in the fiber, $\epsilon$, is proportional to the curvature of the beam, $\kappa$, by
\begin{equation}\label{eq:methods_fbg_euler}
    \epsilon = \kappa y
\end{equation}
where $y$ is the distance from the neutral bending plane of the beam.

\par After eliminating any temperature change effects for \Eqref{eq:methods_fbg_bragg} using the method described in Sec. \ref{sec:methods_tcomp}, from combining Eqs. \eqref{eq:methods_fbg_bragg} and \eqref{eq:methods_fbg_euler} a direct linear proportionality between the shift in FBG's Bragg wavelength and the curvature of the rod is observed by
\begin{equation}\label{eq:methods_fbg_proportional}
    \kappa = \frac{1}{\lambda_{B, 0} S_\epsilon y} \cdot \Delta \lambda_B = c \cdot \Delta \lambda_B,
\end{equation}
where $c$ is a constant of proportionality.
Combining three-channels in different directions and \Eqref{eq:methods_fbg_proportional} of the rod's cross-section, we derive a linear relationship between the curvature induced in the rod's AA, $\boldsymbol{\kappa} = ( \kappa_x, \kappa_y )^T$, and the wavelength shifts of the three different channels in the AA, $\Delta\boldsymbol{\lambda}_B = ( \Delta\lambda_{B,1}, \Delta\lambda_{B,2}, \Delta\lambda_{B,3} )^T$, as
\begin{equation}\label{eq:methods_fbg_calib}
    \boldsymbol{\kappa} = C \cdot \Delta\boldsymbol{\lambda}_B
\end{equation}
where $C \in \reals^{ 2 \times 3 }$ is defined as the constant \emph{calibration matrix} of the AA.

\subsection{Temperature Compensation}\label{sec:methods_tcomp}
\par As mentioned in \Eqref{eq:methods_fbg_bragg}, the Bragg wavelength shift is temperature-dependent \cite{Lee2013, Issatayeva2021}. 
In order to directly compute strain, a method to remove the effect of temperature from the sensor measurements is warranted.
Using the method presented in [\citenum{Cheng2023}], we present the temperature compensation method here.

\par Given that the channels corresponding to an AA are close to each other, we make the assumption that they will experience the same temperature at any point of time.
Furthermore, since the fibers are identical, their temperature coefficients are assumed to be equivalent.
Therefore, wavelength shifts induced in the AA from temperature changes are equivalent between all of the channels in an AA. 
Therefore, we remove the temperature's effect on the wavelength shift by deducting the common mode of the channels' wavelength shifts within an AA \cite{Issatayeva2021, Cheng2023}.

\subsection{Shape Reconstruction Model}\label{sec:methods_shape}
\par Using our sensor-based Lie-group theoretic model, \cite{Kim2017, Lezcano2020, Lezcano2022} we describe the local curvature ($\omega_1$ and $\omega_2$ along the local $x$- and $y$-axes, respectively) and torsion ($\omega_3$ along the local $z$-axis) of the needle as

\begin{figurehere}
    \centering
    \includegraphics[width=0.8\linewidth]{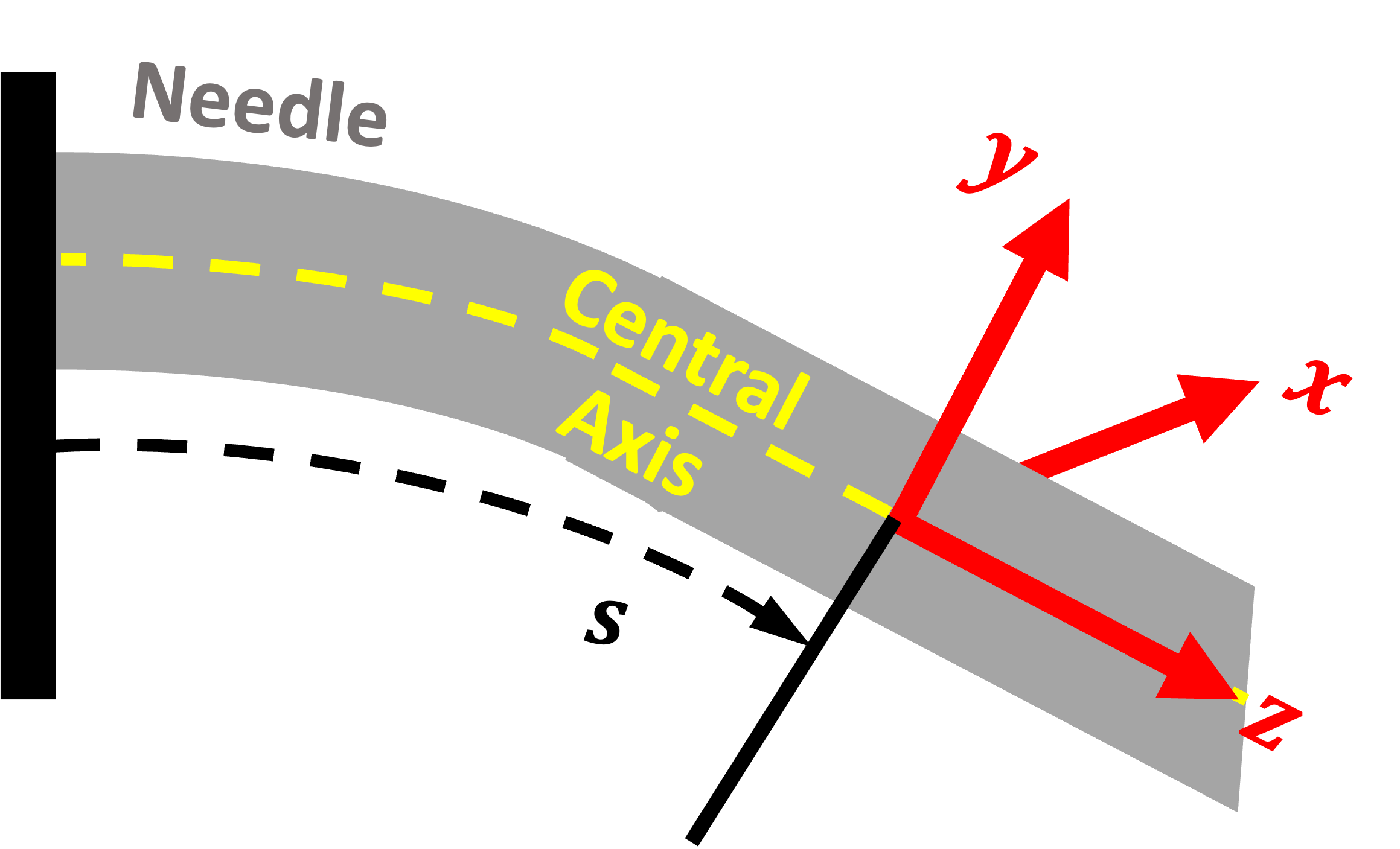}
    \caption{
    The local body-fixed frame of the needle relative to its central axis. 
    }
    \label{fig:methods_shape_axes}
\end{figurehere}
\noindent 

\begin{equation}\label{eq:methods_shape_omegaR}
    {\boldsymbol \omega} (s) = \left[ \omega_1 \,\, \omega_2 \,\, \omega_3 \right]^T = \left( R^T (s) {\textstyle \frac{dR(s)}{ds}} \right)^{\vee}
\end{equation}
where $R(s): \reals \rightarrow SO(3)$ denotes the orientation of the needle's local body-fixed frame, parameterized along the needle's arclength.

The arclength is denoted by $s \in [0, \, L]$ for an insertion with insertion depth $L$. 
The $(\cdot)^\vee: \reals^3 \rightarrow so(3)$ operation defines a function mapping a 3D real-valued vector to the Lie algebra of $SO(3)$, $so(3)$, a set of $3\times 3$ skew-symmetric matrices where $R^T \frac{d R}{d s} \in so(3)$\cite{Chirikjian2016}.

\par For single-layered tissue with a non-rotating insertion of a bevel-tip needle, the needle is modelled as an inextensible elastic rod under uniformly distributed loads \cite{Lezcano2022}.
Naturally, the bevel-tip needle deflects towards the bevel direction, the local $yz$-plane, as show in Fig. \ref{fig:methods_shape_axes}.
Reducing the natural beam mechanics of a uniformly distributed load, we can define an intrinsic curvature that the needle experiences along the needle's arclength as $\kappa_0(s)$, where
\begin{equation}\label{eq:methods_shape_kappa0}
    \kappa_0(s) = \kappac \left(1 - \frac{s}{L} \right)^2.
\end{equation}
Here, $\kappac$ is the intrinsic curvature coefficient, combining the effects of the mechanics of the needle-tissue interaction, to be determined using FBG sensor measurements. 
Ideally, the needle will deflect in the needle's natural bending plane of the needle, in the direction of the needle's bevel-tip. Given this, we denote the 3D intrinsic curvature, $\boldsymbol{\omega}_0(s): \reals \rightarrow \reals^3$, parameterized along the needle's arclength, as 
\begin{equation}\label{eq:methods_shape_omega0}
    \boldsymbol{\omega}_0(s) = \kappa_0(s) \cdot \begin{pmatrix} 1 & 0 & 0\end{pmatrix}^T,
\end{equation} 
using \Eqref{eq:methods_shape_kappa0}.

\par The intrinsic curvature defined in \Eqref{eq:methods_shape_omega0} provides a reference for the needle deformation to follow. Using this, we can define the elastic potential energy of the rod as 
\begin{equation}\label{eq:methods_shape_potential}
    {\mathcal V} = \bigintsss_0^L \frac{1}{2} \left( {\boldsymbol \omega} - {\boldsymbol \omega}_0 \right)^T B \left( {\boldsymbol \omega} - {\boldsymbol \omega}_0 \right) \, ds, 
\end{equation}
where $\boldsymbol{\omega}(s) : \reals \rightarrow \reals^3$, is the local needle deformation parameterized along the needle and $B$ is the needle's stiffness matrix. Minimizing \Eqref{eq:methods_shape_potential} yields the differential equation \cite{Holm1998, Kim2006},
\begin{equation}\label{eq:methods_shape_EPeq}
\frac{d}{ds} \left[ B \left({\boldsymbol \omega} - {\boldsymbol \omega}_0 \right) \right] + {\boldsymbol \omega} \times B \left( {\boldsymbol \omega} - {\boldsymbol \omega}_0 \right) = {\bf 0}
\end{equation}
to be solved in conjuction with \eqref{eq:methods_shape_omegaR} to determine the body-fixed local deformation, $\boldsymbol{\omega}(s)$, and needle orientation, $R(s)$. Finally, the needle shape, $\mathbf{r}(s): \reals \rightarrow \reals^3$, is computed by integrating the body-fixed needle orientation along the length of the needle by
\begin{equation} \label{eq:methods_shape_shape}
    \mathbf{r}(s) = \bigintssss_0^s R(\sigma) \mathbf{e}_3 \, d\sigma, 
\end{equation}
where $\mathbf{e}_3 = \begin{pmatrix} 0 & 0 & 1 \end{pmatrix}^T$

\par Incorporating the FBG sensor measurements of curvature, we can optimize the measured curvature with the model's determined curvature from the solution of \Eqref{eq:methods_shape_EPeq}, by optimizing the model's parameters including the initial angular deformation at the insertion point, $\winit$, and $\kappac$, collectively denoted as ${\boldsymbol \eta}$, using the cost function
\begin{equation}\label{eq:methods_shape_costfn}
{\mathcal C} ({\boldsymbol \eta}) = \sum_{j =1}^m \left\{ \left( {\omega}^{\rm m}_{j,1} - {\omega}_1 (s_j) \right)^2 + \left( {\omega}^{\rm m}_{j,2} - {\omega}_2 (s_j) \right)^2 \right\}
\end{equation}
where $\omega_{j,1}^{\rm m}$ and $\omega_{j,2}^{\rm m}$ denote the FBG's curvature measurements from the $m$-th AA. 
We solve the optimization problem using the interior-point nonlinear optimization algorithm \cite{Nemirovski2008}.

\subsection{Ground Truth Reconstruction: CT and Stereo}
\par Two methods were used for generating ground truth needle shapes for phantom and real-tissue insertion experiments. 
For phantom insertion experiments, the phantom tissue used was transparent and a stereo reconstruction algorithm referred to in [\citenum{Lezcano2022}] was used for 3D needle reconstruction, referenced to have reconstruction errors of $0.16 \pm 0.06 \unit{mm}$.


\par In real tissue, ground truth was generated from CT images after each insertion depth is achieved. 
Since real tissue is not transparent like the phantom tissue, stereo camera visualization is not viable, requiring for another visualization scheme.
A 3D CT scan was used to visualize fiducials and needle inserted into tissue.
The fiducials are used to register the needle's coordinate frame with the CT coordinate frame. 
Fiducials are segmented and localized in the CT image using the k-means algorithm\cite{Hartigan1979}.
After the fiducials locations in CT were determined, a point cloud registration was used to determine the CT coordinate frame relative to the needle's frame.
This registration was used to compare the CT-reconstructed needle shape with needle shape-sensing results.
Then, the needle was segmented from the CT scan using simple thresholding and interpolated with second order, 3D B-splines.
The interpolation allowed for determining a curve of the needle shape with the discrete slices attained from the CT scan.
Second order 3D B-splines were used for their robust ability to fit complex curves \cite{Pham1989}.
CT reconstruction errors were found to be within $0.14 \pm 0.03 \unit{mm}$ using this method.

\section{Experimental Setup}\label{sec:experiments}

\par The two sensorized needles underwent characterization and calibration as seen in [\citenum{Cheng2023}]. After characterization and calibration, needle insertions were performed in gel phantom and real meat for each of the needles. 

\subsection{Characterization}\label{sec:experiments_character}
\par In order to ensure proper construction of the of the needles in Sec. \ref{sec:needles}, we deflected each needle's tip in increments of 1.5 mm to 15 mm at three different loading angles. 
Proper construction of the needles would be indicated by a linear relationship between the needle tip's deflection distance and the wavelength shift observed from the straight configuration.
Characterization was performed as referenced in [\citenum{Song2021,Cheng2023}].
The needles were deflected using a robotic platform similar to the one in Sec. \ref{sec:experiments_phantom} for accurate measurements of the tip deflection. 

\subsection{Calibration and Validation}\label{sec:experiments_calib}

\begin{figurehere}
\begin{minipage}{\linewidth}
    \centering
    \includegraphics[width=\linewidth]{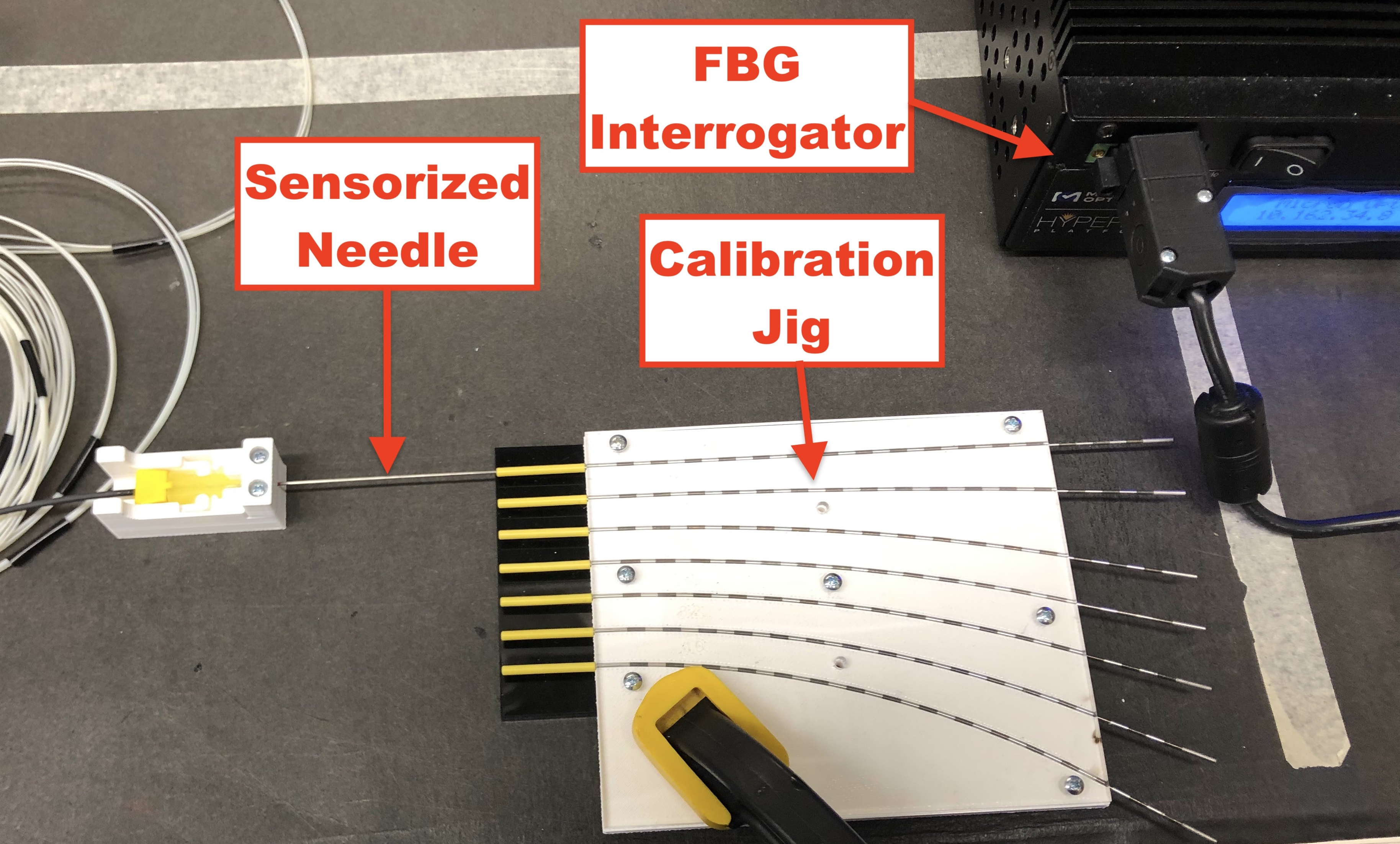}
    \caption{Experimental setup for calibrating FBG-sensorized needles with constant curvature jigs. FBG sensor data was collected using interrogator when the sensorized needle was inserted into the tubes embedded into the constant curvature grooves.}
    \label{fig:experiments_calib_setup}
\end{minipage}
\end{figurehere}

\par After a successful characterization, the FBG-sensorized needles were calibrated with constant curvature jigs in order to determine the relationship referenced in \Eqref{eq:methods_fbg_calib}. 
We designed jigs with constant curvature grooves of varying known curvatures, embedded with larger needle sheaths for the insertion of the sensorized needle.
The constant curvature groove ranged from 0.5 1/m to 4.0 1/m in value, where the jig is shown in Fig. \ref{fig:experiments_calib_setup}.
A straight groove was added to the jigs in order to establish a baseline of unstrained wavelength from all of the FBG sensors for the calculation of wavelength shifts.
Five insertion trials per curved groove at four different needle orientations ($0\degrees$, $90\degrees$, $180\degrees$, and $270\degrees$) were performed to remove any experimental noise from the insertions and to calibrate the sensors for 2D curvature estimation.
Curvatures were distributed into a calibration and validation dataset in order to validate the proper calibration of the needle, using the validation dataset.
For each of the grooves, wavelength shifts were averaged over the five trials to establish which was then used for a linear regression between the 2D curvature experienced by the needle (in the needle's frame) to the averaged wavelength shift experienced by the FBG sensors per AA. 
At the end of the calibration, calibration matrices were derived for each of the AAs, used for 2D curvature estimates at each of the AA locations along the needle.
Finally, using the derived calibration matrices, reliability weightings for all of the active areas were determined as performed in [\citenum{Lezcano2022}] through a linear least squares optimization of the weighted mean-squared error of curvatures across the active areas in order to optimally weight the curvature information provided to the sensor-based shape-sensing method, as FBG sensors are observed to perform differently along the needle.

\subsection{Needle Insertion Robotic Platform}
\begin{figure*}[t!]
    \centering
    \includegraphics[width=0.8\linewidth]{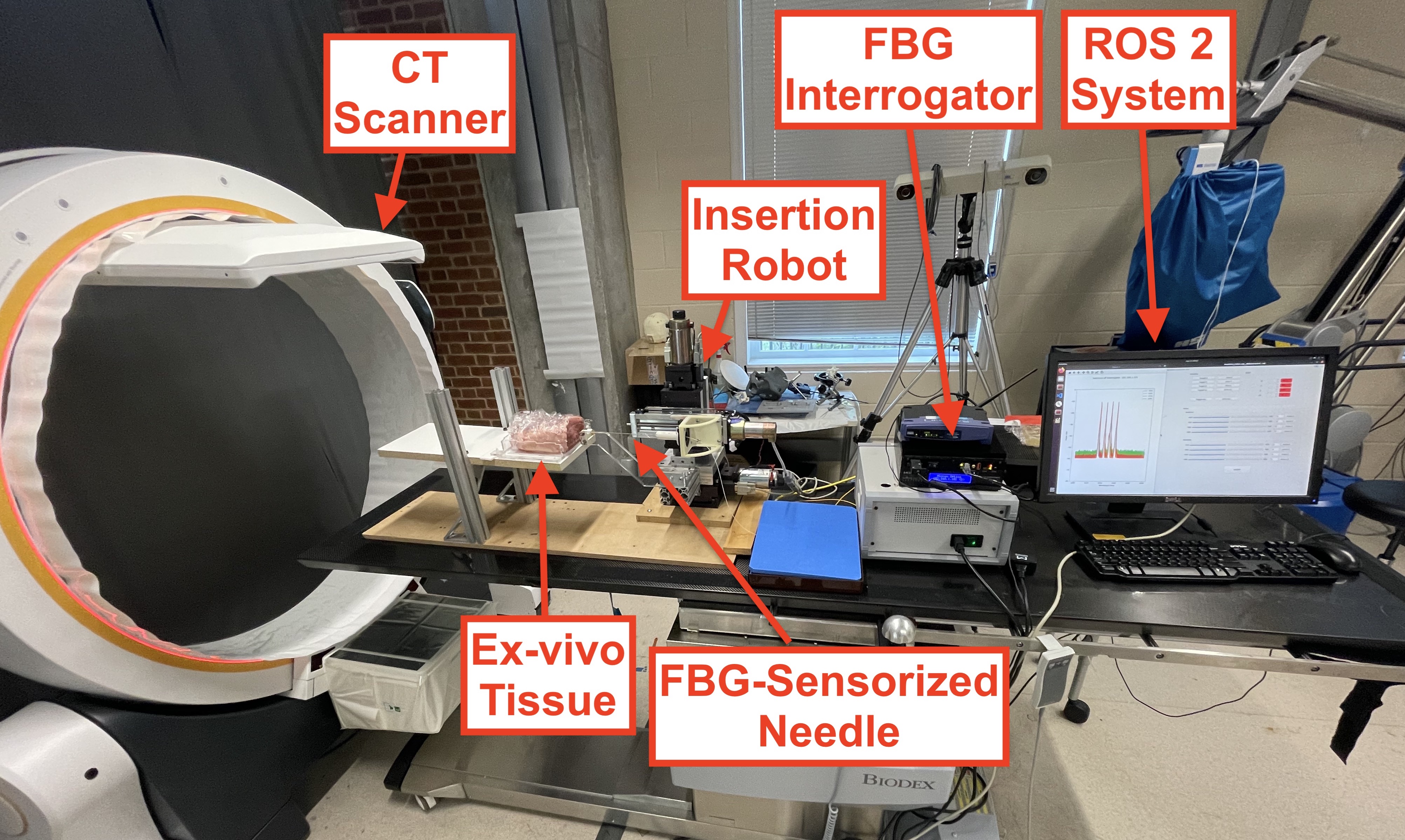}
    \caption{Robotic insertion experimental setup for \exvivo needle insertions. Needle insertion was performed using a needle insertion robot with 4 DoFs, controlled with a ROS 2 system. FBG sensor feedback was streamed over the ROS 2 network, where real-time needle shape-sensing results were provided. The needle's shape was visualized using a Loop-X CT scanner.}
    \label{fig:experiments_ct_setup}
\end{figure*}
\par For both gel phantom and \exvivo tissue insertion experiments, the same robotic insertion platform was used.
The only difference between the two experiments was the modality for ground truth needle shape generation.
To hold the tissue subject in-place, an acrylic box of dimensions allowing for 130 mm insertion depths was used during needle insertion experiments.
For gathering FBG sensor data from either sensorized needle, an optical interrogator (HYPERION si155, Luna Inc., Virginia, United States) was used to collect 200 FBG wavelength samples, for each insertion depth per trial.
Unstrained wavelengths are collected prior to the start of each insertion experiment to establish a baseline for calculating wavelength shifts.
Needle insertion was performed using a 4-degree of freedom (DoF) robotic insertion platform integrated into a ROS 2 system \cite{ROS2}. 
The 4-DoF robotic insertion platform was identical to the one in [\citenum{Lezcano2022}].
The robotic insertion platform was attached with a manual rotation stage containing a custom 3D-printed needle holder.

\par The ROS 2 system contained custom-written packages for handling the ground truth stereo vision, for phantom insertion experiments; FBG interrogator nodes; needle shape-sensing; and robotic control.
A custom user interface was developed for robotic platform control and RQT was used for needle visualization.
The ROS bagging system was used for recording data to be post-processed after each experiment.

\subsection{Phantom Insertions}\label{sec:experiments_phantom}
\par 
For SCF needle insertion into gel phantom, experimental results were used from [\citenum{Lezcano2022}] since the SCF needle configurations were identical between that work and this one, as well as similar calibration performance, as described in Sec. \ref{sec:results_calib}. 
The MCF needle was inserted into soft gel for five trials at insertion depths of 30, 60, 90, and 120 mm.

\par The homogeneous gel phantom was constructed according to [\citenum{Lezcano2022}]. 
Plastic (MF-Manufacturing Company, Texas, United States) melted and formed into a mold for creating the phantom tissue.

\subsection{Ex-vivo Tissue Insertions}\label{sec:experiments_realtissue}
\par 
Needle insertion into \exvivo tissue was performed for both the SCF and MCF needles for insertion depths of 65 and 125 mm.
The SCF needle was inserted into pork tissue for three trials and the MCF needle was inserted into beef tissue for 5 trials.
Nine registration fiducials were attached to the acrylic box to identify the coordinate system of the CT scanner.
A CT scanner (Loop-X, Brainlab, Munich, Germany) visualized the needle inserted into tissue as well as fiducials attached to the acrylic box, holding the tissue, shown in Fig. \ref{fig:experiments_ct_setup}.
Each CT image was a collection of 2D slices with pixel spacings of 0.447 mm/pixel, and each slice thickness was 0.667 mm, with a field of view with dimensions approximately of 20 cm $\times$ 25 cm $\times$ 23 cm.
Needle insertion trials were limited to the CT scanner's available number of 3D scans prior to overheating, hence the fewer number of performed insertion trials than in phantom tissue.

\section{Results}\label{sec:results}
\par We used several metrics to compare the ground truth needle shape, $\mathbf{r}_{gt}$, to the sensed needle shape, $\mathbf{r}$, that are discretized by their arclength $s_i,\, i = 1, ..., N$. They are listed below:

\par \textbf{Tip Error (TE)}: the location error from the tip of the needle.
\begin{equation}\label{eq:results_error_tip}
    TE = \norm{\mathbf{r}_{gt}(L) - \mathbf{r}(L)}
\end{equation}

\par \textbf{Root-Mean Square Error (RMSE)}: the overall RMS error of the needle shape.
\begin{equation}\label{eq:results_rmse}
    RMSE = \sqrt{
        \frac{1}{N} \sum_{i=1}^N 
        \norm{\mathbf{r}_{gt}(s_i) - \mathbf{r}(s_i)}^2
    }
\end{equation}

\par \textbf{In-Plane Error (IPE)}: the error measured in the natural bending plane of the needle.
\begin{equation}\label{eq:results_error_inplane}
    IPE = \frac{1}{N} \sum_{i=1}^N \norm{
        \begin{pmatrix} 0 & 1 & 1 \end{pmatrix}
        \cdot \left( \mathbf{r}_{gt}(s_i) - \mathbf{r}(s_i)  \right)
    }
\end{equation}

\par \textbf{Out-of-Plane Error (OPE)}: the error measured in the plane orthogonal to the natural bending plane of the needle.
\begin{equation}\label{eq:results_error_outplane}
    OPE = \frac{1}{N} \sum_{i=1}^N \norm{
        \begin{pmatrix} 1 & 0 & 1 \end{pmatrix}
        \cdot \left( \mathbf{r}_{gt}(s_i) - \mathbf{r}(s_i)  \right)
    }
\end{equation}

\par \textbf{Max Error (MAX)}: the maximum error measured along the needle.
\begin{equation}
    MAX = \max_{i} \left(
        \norm{\mathbf{r}_{gt}(s_i) - \mathbf{r}(s_i)}
    \right)
\end{equation}

Note that $\cdot$ here denotes matrix multiplication.

\subsection{\nameref{sec:experiments_character}}\label{sec:results_character}

\begin{figurehere}
    \centering
    \begin{subfigure}[b]{\linewidth}
        \centering
        \includegraphics[width=\linewidth]{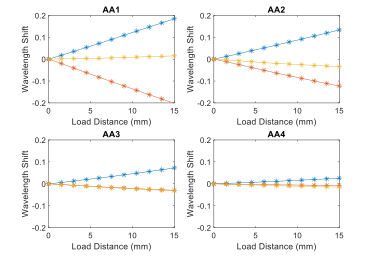}
        \caption{SCF needle}
    \end{subfigure}
    \begin{subfigure}[b]{\linewidth}
        \centering
        \includegraphics[width=\linewidth]{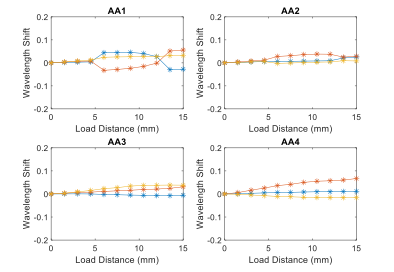}
        \caption{MCF needle}
    \end{subfigure}
    \caption{Characterization results from [\citenum{Cheng2023}] for the SCF and MCF needle over a loading distance of 15 mm in 1.5 mm increments.}
    \label{fig:results_character}
\end{figurehere}

\par Characterization results presented were from [\citenum{Cheng2023}].
The SCF needle presented a linear relationship with the loading of the needle's tip, in all of the channels. 
We also observed that AA1, furthest from the needle tip, was strained the most with the largest wavelength shifts, while AA4, closest to the needle tip, was strained the least with the smallest wavelength shifts. 
The characterization results indicated a proper construction enabling a proper calibration of these sensors.

\par The MCF needle demonstrated a linear relationship within a small-loading regime and afterwards reached a second regime where there was a jump into different linear regime. 
Particularly seen in AA1, we observed a shift in the slope of the wavelength response to the loading distance. 
There was also a jump where AA1 became the most activated. 
Similar to the SCF, in the first linear regime AA4, furthest from the tip, experienced the most strain induced in the FBGs, while AA1, experienced the least.
The MCF needle in this construction experienced a large amount of non-linearity, increasing in non-linearity as the FBG sensor approached the tip. 
Notably, the scale of the signal response experienced by the MCF FBGs was much smaller than the ones induced in the SCF. 
This feature could indicate a potential low signal-to-noise ratio (SNR) for the MCF, potentientially attributing to the non-linear behavior observed by the MCF characterization.

\subsection{\nameref{sec:experiments_calib}}\label{sec:results_calib}

\par To justify using the SCF needle shape-sensing results from [\citenum{Lezcano2022}], we use a t-test to compare that needle's calibration with this work's. 
Using a two-tailed t-test to compare SCF from [\citenum{Lezcano2022}] and this work yielded presented in Table \ref{tab:results_calib_comp}: 
AA1 \textemdash\, $\abs{t} = 0.31 < 1$, 
AA2 \textemdash\, $\abs{t} = 0.56 < 1$, 
AA3 \textemdash\, $\abs{t} = 0.62 < 1$, 
AA4 \textemdash\, $\abs{t} = 0.26 < 1$, 
found to be insignificant, concluding that the SCF needles perform similarly.

\begin{tablehere}
\begin{minipage}{\linewidth}
    \centering
    \caption{Calibration curvature error statistics, mean and standard deviation, comparing the SCF and MCF needles from this work and [\citenum{Lezcano2022}] for each active area.}
    \begin{tabular}{|c|c|c|c|}
        \hline
         \multirow{2}{*}{\textbf{AA}} &  \multicolumn{3}{c|}{\textbf{Curvature Error (1/m)}}  \\\cline{2-4}
            &  \textbf{[\citenum{Lezcano2022}] SCF} & \textbf{This SCF} & \textbf{MCF} \\  
         \hline
         1 & $0.16 \pm 0.11$ & $0.21 \pm 0.12$ & $0.35 \pm 0.27$ \\
         2 & $0.15 \pm 0.11$ & $0.08 \pm 0.06$ & $0.50 \pm 0.34$ \\
         3 & $0.34 \pm 0.24$ & $0.15 \pm 0.10$ & $0.31 \pm 0.28$ \\
         4 & $0.47 \pm 0.39$ & $0.36 \pm 0.16$ & $0.55 \pm 0.44$ \\
         \hline
    \end{tabular}
    \label{tab:results_calib_comp}
\end{minipage}
\end{tablehere}

\par A notable feature found in Table \ref{tab:results_calib_comp} is that the curvature estimation error was much higher and less precise in the MCF needle than the SCF needle. 
This indicates an issue when trying to reconstruct the shape as the calibration was not as reliable. 
The increased error from the MCF could be attributed to the low SNR found in Fig. \ref{fig:results_stats_mcf_real}.

\subsection{\nameref{sec:experiments_phantom}}\label{sec:results_phantom}
\par
\begin{figure*}
    \centering
     \begin{subfigure}[b]{0.45\linewidth}
         \centering
         \includegraphics[height=6cm, width=\textwidth]{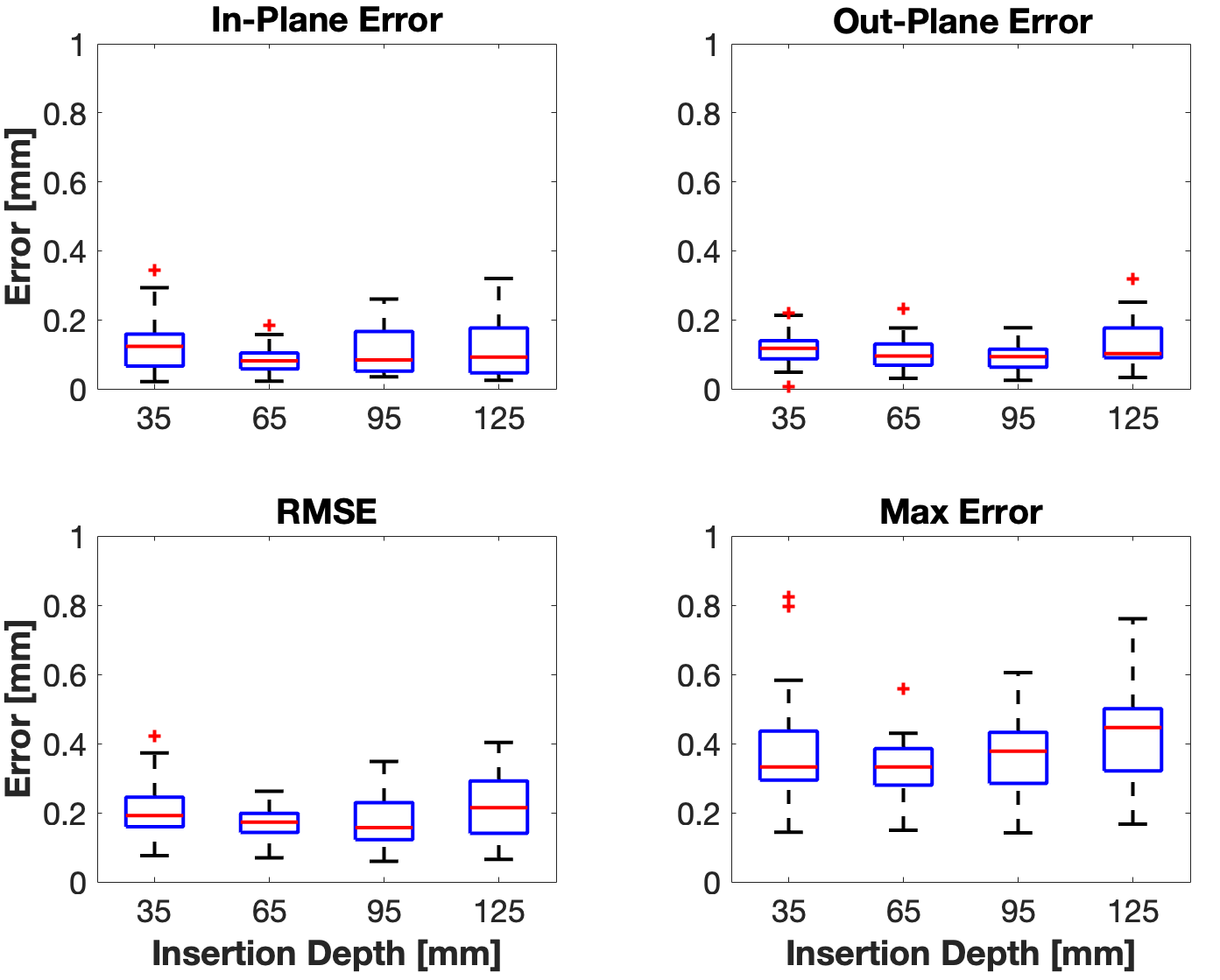}
         \caption{SCF in phantom}
         \label{fig:results_stats_scf_phantom}
    \end{subfigure}
    \begin{subfigure}[b]{0.45\linewidth}
         \centering
         \includegraphics[height=6cm, width=\textwidth]{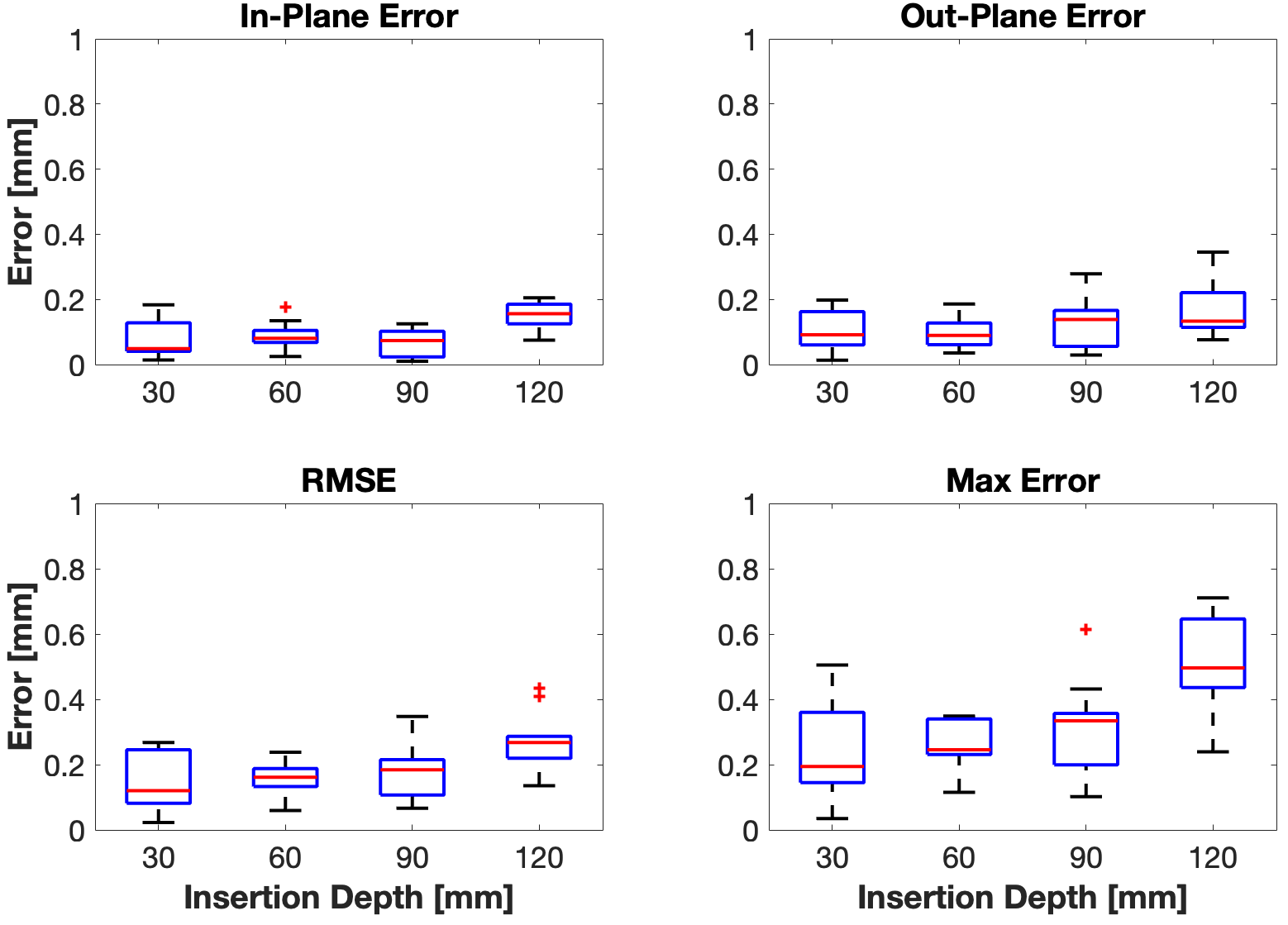}
         \caption{MCF in phantom}
         \label{fig:results_stats_mcf_phantom}
    \end{subfigure}
    \begin{subfigure}[b]{0.45\linewidth}
         \centering
         \includegraphics[height=6cm, width=\textwidth]{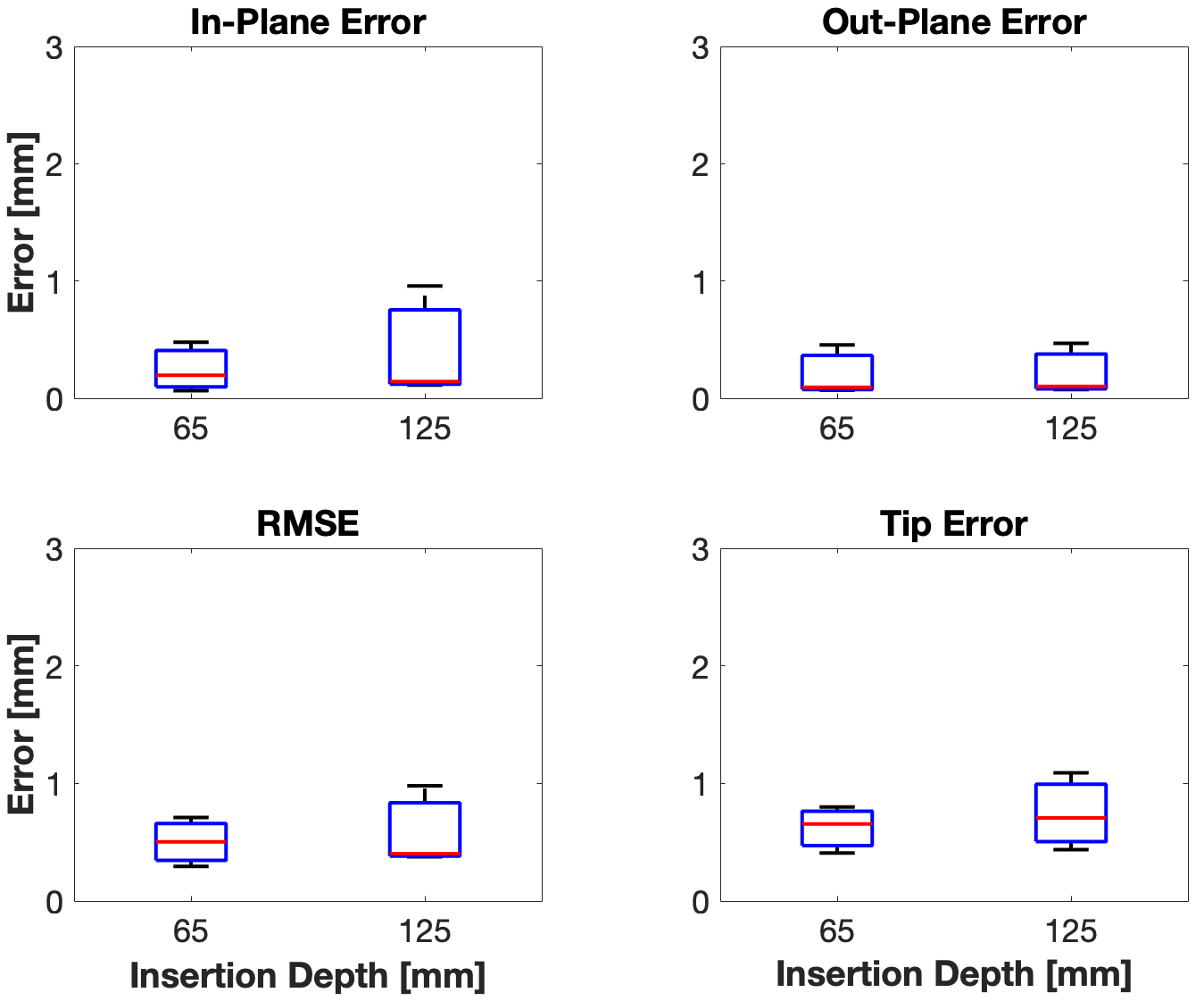}
         \caption{SCF in real tissue}
         \label{fig:results_stats_scf_real}
    \end{subfigure}
    \begin{subfigure}[b]{0.45\linewidth}
         \centering
         \includegraphics[height=6cm, width=\textwidth]{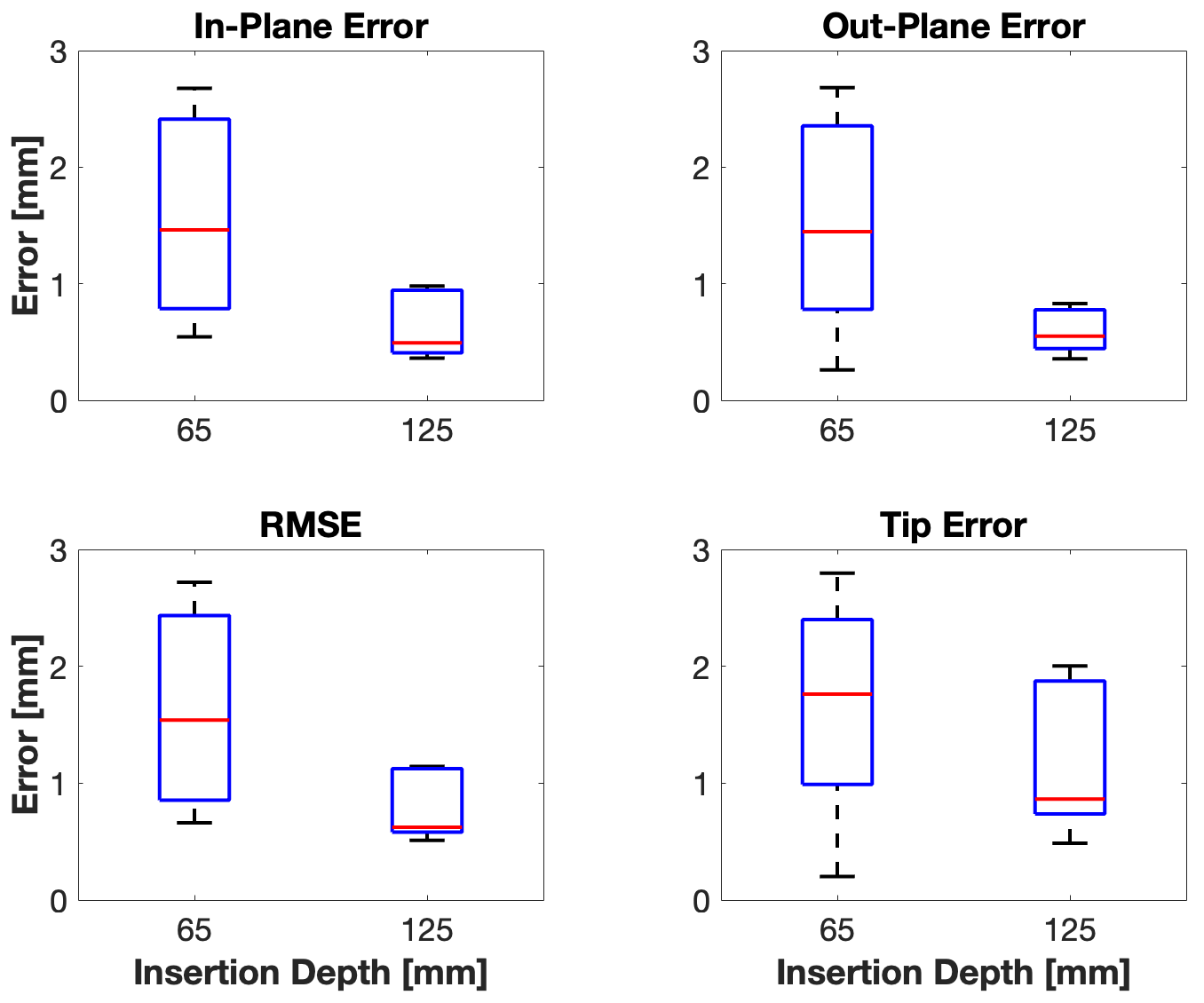}
         \caption{MCF in real tissue}
         \label{fig:results_stats_mcf_real}
    \end{subfigure}
    \caption{Shape-sensing error statistics for the SCF and MCF needles in phantom, (\subref{fig:results_stats_scf_phantom}) and (\subref{fig:results_stats_mcf_phantom}), real, (\subref{fig:results_stats_scf_real}) and (\subref{fig:results_stats_mcf_real}), tissue. 
    Ground truth needle shapes were measured from stereo and 3D CT reconstruction for the phantom and real tissue insertions, respectively.}
    \label{fig:results_stats}
\end{figure*}

 
Insertion experiment results are presented for the soft-tissue single-layer C-shape insertion for the MCF-sensorized needle for insertion depths of 30, 60, 90, and 120 mm, and from [\citenum{Lezcano2022}] for the SCF needle at comparable insertion depths of 35, 65, 95, and 125 mm.
Demonstrated in Figs. \ref{fig:results_stats_scf_phantom} and \ref{fig:results_stats_mcf_phantom}, we see that all shape-sensing errors for both the SCF- and MCF-sensorized needles were within 1 mm, with average errors within 0.5 mm. 
When looking at the error contributions between the IPE and OPE to the total RMSE, we observed equal contributions, indicating uniform performance of the SCF and MCF sensors in the needles in gel phantom.
Notably in Fig. \ref{fig:results_stats_mcf_phantom} as compared to Fig. \ref{fig:results_stats_scf_phantom}, the MCF needle at the maximum insertion depth had a spike for shape-sensing error, as compared to the SCF needle where shape-sensing error remained comparable. 
Both the SCF and MCF needle overall performed comparably with each other, yielding errors that were well within each other's range.
When performing a $p$-test between the SCF and MCF shape-sensing results in phantom tissue, we get a $p = 0.164 > 0.05$ for the RMSE, indicating insignificant discrepancy between the two needles' performances.
Overall shape-sensing errors for these insertion depths were $0.35 \pm 0.12 \unit{mm}$ and $0.19 \pm 0.09 \unit{mm}$ for the SCF needle and MCF needle, respectively.


\subsection{\nameref{sec:experiments_realtissue}}\label{sec:results_realtissue}

\par Insertion experiments in \exvivo tissue are reported for the SCF and MCF needle insertions for insertion depths of 65 and 125 mm. 
The SCF needle presented similar results in real tissue as compared to in phantom tissue. 
All errors were within 1 mm, with average shape-sensing errors hovering around 0.5 mm. 
Furthermore, we see that the shape-sensing errors remained consistent between varying insertion depths. 
We observed large insertion errors, up to 2.5 mm, for the MCF needle, on the other hand.
Particularly, this error spiked for smaller insertion depths, when only two AAs were inserted into the tissue. 
At full insertion depth, the MCF needle performed similarly to its performance in phantom tissue with larger averages. 
However, we observed a large error at the tip from the MCF needle at all insertion depths, ranging up to 2 mm for both insertion depths.
$p$-value between SCF and MCF yield for RMSE: $p = 0.0005 < 0.05$, which indicated a significant discrepancy. 
Overall shape-sensing errors at these insertion depths were $0.64 \pm 0.31 \unit{mm}$ and $1.33 \pm 0.65 \unit{mm}$ for the SCF needle and MCF needle, respectively. 

\section{Discussion}\label{sec:discussion}
\par For phantom insertions, we observed that both the SCF- and MCF-sensorized needles perform similarly. 
\begin{figure*}[t]
    \centering
    \includegraphics[width=0.9\linewidth]{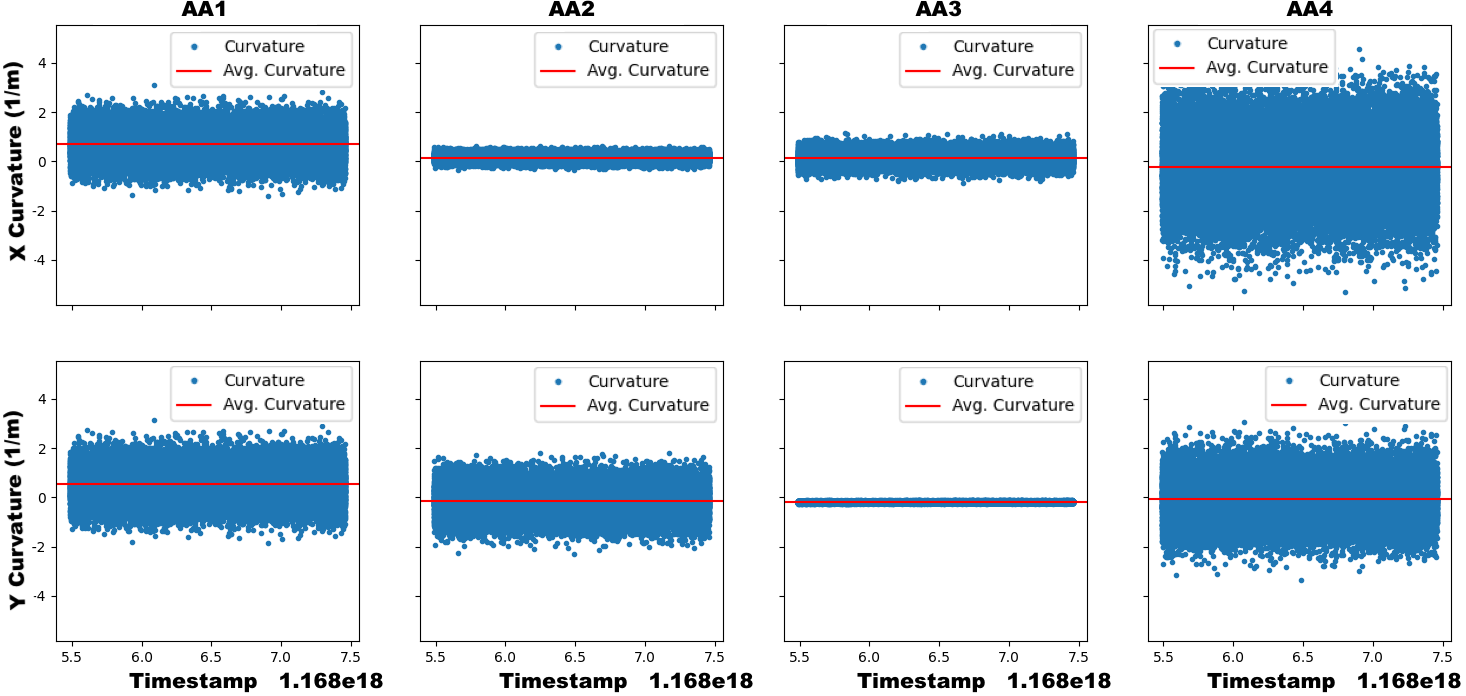}
    \caption{The $x$- (top) and $y$-curvature (bottom) sensed in the MCF needle's AAs over time during a static insertion depth of 125 mm in \exvivo tissue. The sensor noise in AA1 and AA4 dominate the signal's output and contribute to large error in the measured curvature.}
    \label{fig:discussion_sensornoise}
\end{figure*}
As mentioned in Sec. \ref{sec:results_phantom}, the calculated $p$-value indicated insignificant discrepancy between the shape-sensing error distributions for these insertions. 
Furthermore, average errors were within 0.5 mm with the exception of the MCF needle at 125 mm insertion depth.
At 125 mm insertion depth of the MCF, we observed a jump in the maximum error measured in the FBG-based shape-sensing. 
This could be attributed to the noise found in AA4. 
Upon inserting from 125 mm from 95 mm, AA4 was inserted into the tissue and then was used for shape-sensing.
As found in Fig. \ref{fig:discussion_sensornoise}, AA1 and AA4 experienced large amounts of sensor noise affecting the curvature reconstruction of the needle.
This was largely due to the strain induced in these fibers was not large enough in order to strain.
Furthermore, the construction of the MCF needle was embedded with the MCF by gluing the base of the needle to the sensor, holding the sensor as taut as possible.
However, AA4 was located at the middle of the 200 mm length needle, where the sensor experienced significant slack, therefore reducing the needle's strain transfer to this sensor.
As seen in Fig. \ref{fig:discussion_sensornoise}, AA4's curvature estimation was found to be entirely noisy and was deemed unreliable for needle shape estimation. 
Due to this sensor noise, AA4 of the MCF was manually given a minimal reliability weight in order to remove it from needle shape estimation.
Due to loose tolerances in the constant curvature jig, noise and error are experienced when calibrating the FBG sensors closest to the tip of the needle.
Therefore, the noise seen in Fig. \ref{fig:discussion_sensornoise} was largely contributed to calibration errors. 
These tolerances could potentially be mitigated through tightening the tolerances of the constant curvature jig. However by reducing the inner diameter of the tubes used to calibrate the needles, it becomes very difficult to insert the sensorized needle into the jig.
With this difficulty, large forces and manipulation are required to insert the needle into the jig, increasing the risk of breaking the needle and embedded sensors. 
Furthermore, a finite element model could be used to estimate what the experienced curvature by incorporating the estimated tolerances of the embedded sensors in the MCF, however we leave this as a future work.

\par 
As seen in Fig. \ref{fig:discussion_maxdeflection}, the MCF needle has the largest shape-sensing error when the needle was deflected less.
All of the shape-sensing errors were found to be when the maximum deflection was less than 6 mm, primarily seen in the real-tissue insertion experiment. The average maximum deflection observed over all of the MCF needle insertion trials 

\begin{figurehere}
    \centering
    \includegraphics[width=\linewidth]{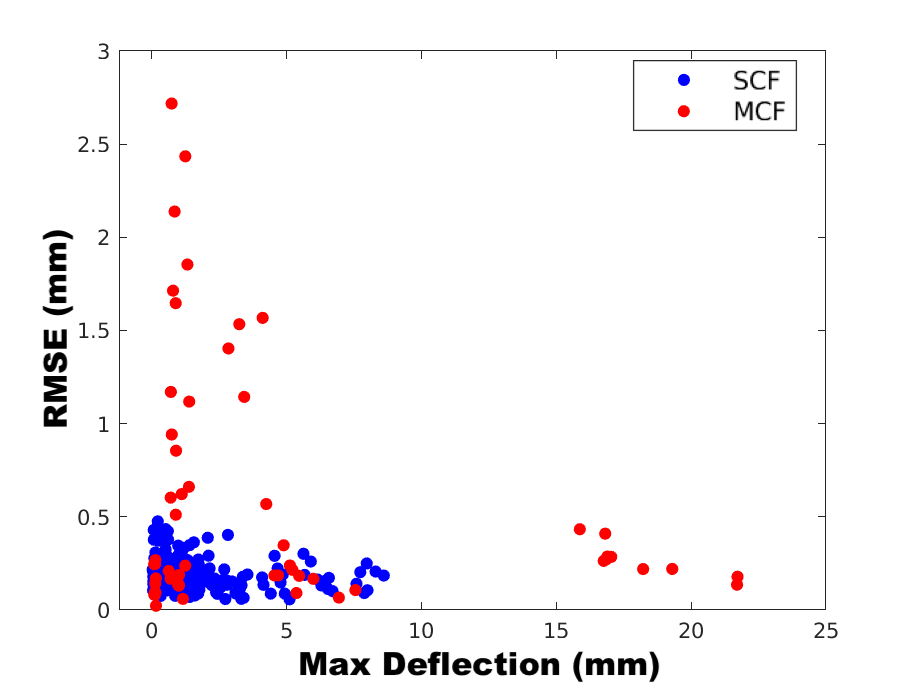}
    \caption{The relationship between the needle's sensed maximum deflection to shape-sensing error (RMSE) found for the SCF and MCF sensors embedded into flexible needles over all insertion depths.}
    \label{fig:discussion_maxdeflection}
\end{figurehere}
\noindent
for the maximal insertion depth in soft-tissue gel phantom was $18.1 \pm 2.1 \unit{mm}$, while in real tissue was $2.4 \pm 1.4 \unit{mm}$.
The larger shape-sensing errors were expected to occur at smaller deflections of the needle since the embedded MCF sensors in this construction suffered from low SNR. 
Thus, when the needle was deflected less, the FBGs experienced less strain, exasperating the SNR issue and finally deteriorating shape-sensing performance.
Compared to the SCF needle, while maximum deflections were similar to those found in the MCF needle in \exvivo tissue, the configuration of the fibers in this needle did not suffer from the low SNR problem found in the MCF needle's configuration.
Since the fibers in the SCF needle were embedded radially further from the needle's central axis, the strain transferred to the SCF from needle bending was much larger than found in the MCF, as the MCF was co-axially mounted with its needle.
This is justified by looking at the needle's characterization results found in Sec. \ref{sec:results_character}, as the SCF needle experienced larger and more linear signal responses over all of the AAs, as compared to the MCF. 
Furthermore, since adhesive was uniformly applied to the fibers in the SCF needle, the fibers were consistently joined to the needle, while in the MCF needle, there was only one mounting point, causing for the extra noise found from AA4 in the MCF needle.

\par From this study, we find that single-core fibers and multicore fibers have distinct advantages and disadvantages for needle shape-sensing.
Using SCFs, we found that controlling the direct placement of each of the sensing benefited needle shape-sensing performance since strain transfer is able to be maximized by placing the sensors further from the needle's central axis.
Since needles are typically not bent in large curvatures like endoscopes or catheters, fine realization of curvatures for small curvatures is imperative for proper shape-sensing performance. 
Thus, since MCF FBGs are constrained by their closer placement within the sensing array, the FBGs across the array experiences a limited variation of strain transfer.
This limited variation of strain transfer over the MCF's cross-section warrants a lack of fine estimation of needle curvature, especially when the MCF is co-axially mounted with the needle, as the MCF is minimally strained in this configuration.
However, using the MCF's central core with a co-axially mounted MCF, temperature compensation is able to be performed since the MCF's central core undergoes no strain, while in SCF fibers, the temperature compensation method presented in Sec. \ref{sec:methods_tcomp} works empirically.
Furthermore, given that the MCF is mounted with seven FBGs, the MCF sensor configuration is able to determine its own shape without relying upon mechanical models.
However, this would require a more dense placement of FBGs along the sensor or distributed FBGs, hence the necessity of using the shape-sensing model presented in this paper.
For SCFs, a medium is required to attach multiple fibers to enable 3D shape-sensing, as the SCFs alone are unable to determine their own shape. 
Thus, SCF calibration is on a per-needle basis, while MCFs can be interchanged between needles given that the MCF location is constrained mechanically to a single cable.
Embedding SCFs into needles is also more intensive and laborous task with many points of failure, while MCF embeddings has a single point of failure. 
While embedding SCFs, we have encountered issues with improper adhesion to the needle as the SCF needs full adhesion to the needle, as well as potential twisting of SCFs inside the inner stylet's grooves. 
However, given that the MCFs are only mounted at the base of the needle, strain transfer at points closer to the center of the needle degrade due to slack of the sensor inside the needle, as seen in Fig. \ref{fig:discussion_sensornoise}.
Finally, fiber adhesion degrades over many uses of the needle, and since SCF embedded needles rely heavily on proper adhesion of the fibers, SCFs will need to be re-glued to the needle and the needle will need to be recalibrated for research purposes. 
In contrast, as MCF embedded needles degrade, the only requirement to fix the needle is to re-apply glue to at one, accessible point on the needle and does not require recalibration since the FBG locations inside the MCF relative to each other is fixed.
In practice, these needles will be treated as a disposable consumable, where they will be discarded once the needle deviates from its specified calibration.
Therefore, the cost of MCFs become a larger concern to their viability, however, with only one point of failure, it is likely the MCF-sensorized needles will fail after more insertions than the SCF-sensorized needles.

\par For our current usage of MCFs in this comparative study, we found better shape-sensing reliability with SCFs. 
Nonetheless, we still find MCF-based needle shape-sensing as an important research for continued study, thus we provide points of potential improvement that could be used for better using MCFs in needles.
Firstly, using a higher resolution interrogator to better resolve MCF signals for small curvature estimation would mitigate the encountered SNR issue.
An off-axis placement of the MCFs could be used to increase strain transfer to the FBGs, further addressing the MCF's low SNR.
In this study, we only used four of the MCF's channels, the central core and three outer cores, due to hardware limitations of our four-channel interrogator.
To use all of the cores with standard market interrogators, an interrogator with at least eight channels are required which greatly increases the cost of using these sensors in needle shape-sensing, thus we proceeded with our four-channel interrogator to directly compare their shape-sensing performance in a cost-effective manner.
Better MCF performance could be attained through using all of the MCF's outer cores, which may also balance the low SNR through using redundant sensors.
Moreover, methods to address inherent twist in the MCFs were not used in this study in order to directly compare the raw shape-sensing performance of these sensors, though these methods may increase the MCF's performance to realize the needle's 3D shape.

\section{Conclusion}\label{sec:conclusion}
\par This paper provides a baseline evaluation of SCFs and MCFs performance for needle shape-sensing through identical experiments, establishes a method for evaluating future optical sensors for needle shape-sensing, and provides points of improvement for integrating MCFs into needles for shape-sensing tasks.
We configured these sensors identically in identical needles in order to provide a direct comparison of raw shape-sensing capabilities in needles, with similar costs to fabricate and use these needles.
We realized mean accuracies for SCF-based needle shape-sensing of $0.35 \pm 0.13 \unit{mm}$ and $0.64 \pm 0.31 \unit{mm}$ for phantom and \exvivo tissues, respectively.
MCF-based needle shape-sensing performance was found to have average accuracies of $0.19 \pm 0.09 \unit{mm}$ and $1.33 \pm 0.65 \unit{mm}$ for phantom and \exvivo tissues, respectively.
We found that MCF-based shape-sensing in phantom tissue, where the needle incurred the largest deflection, performed similarly to the SCF-based configuration with a $p$-value of $0.164 > 0.5$, but in \exvivo tissue the MCF needle performed drastically worse than the SCF needle with a $p$-value of $0.0005 < 0.5$ due to low SNR found in the MCF.
Points of improvement for MCF-based needle shape-sensing are provided in this paper to mitigate the low SNR issue found in MCF-embedded needles.
Limitations of this work include using only four out of the seven channels in the MCF needle, the smaller sample size of experimental insertions in \exvivo tissue, and the lack of testing MCF twist compensation methods in order to directly compare raw capabilities of the SCF and MCF sensor for a baseline evaluation.
Future work includes extending this study to the utilization of all MCF channels for needle curvature estimation, evaluation of distributed sensing modalities as compared to discrete sensing, and implementing and testing the points of improvement for MCF-based needle shape-sensing presented in this paper for MCF constructive optimization.

\nonumsection{Acknowledgments}
\noindent This work was supported by the National Institutes of Health under grant No.~R01CA235134 and in part by Johns Hopkins University internal funds.

\bibliographystyle{ws-jmrr}
\bibliography{references.bib}

\noindent\includegraphics[width=1in]{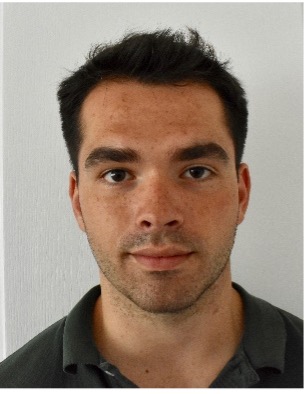}
{\bf Dimitri A. Lezcano} received the M.S.E in Robotics from Johns Hopkins University, Baltimore, Maryland, USA and the B.A. degree in Physics and Mathematics from McDaniel College, Westminster, Maryland, USA, in 2019 and 2021, respectively. He is currently pursuing the Ph.D degree in Mechanical Engineering from Johns Hopkins University, Baltimore, Maryland, USA.\\
His research interests are in computer-integrated surgical technologies, including flexible needle shape-sensing, needle trajectory generation, and flexible needle steering.

\noindent\includegraphics[width=1in]{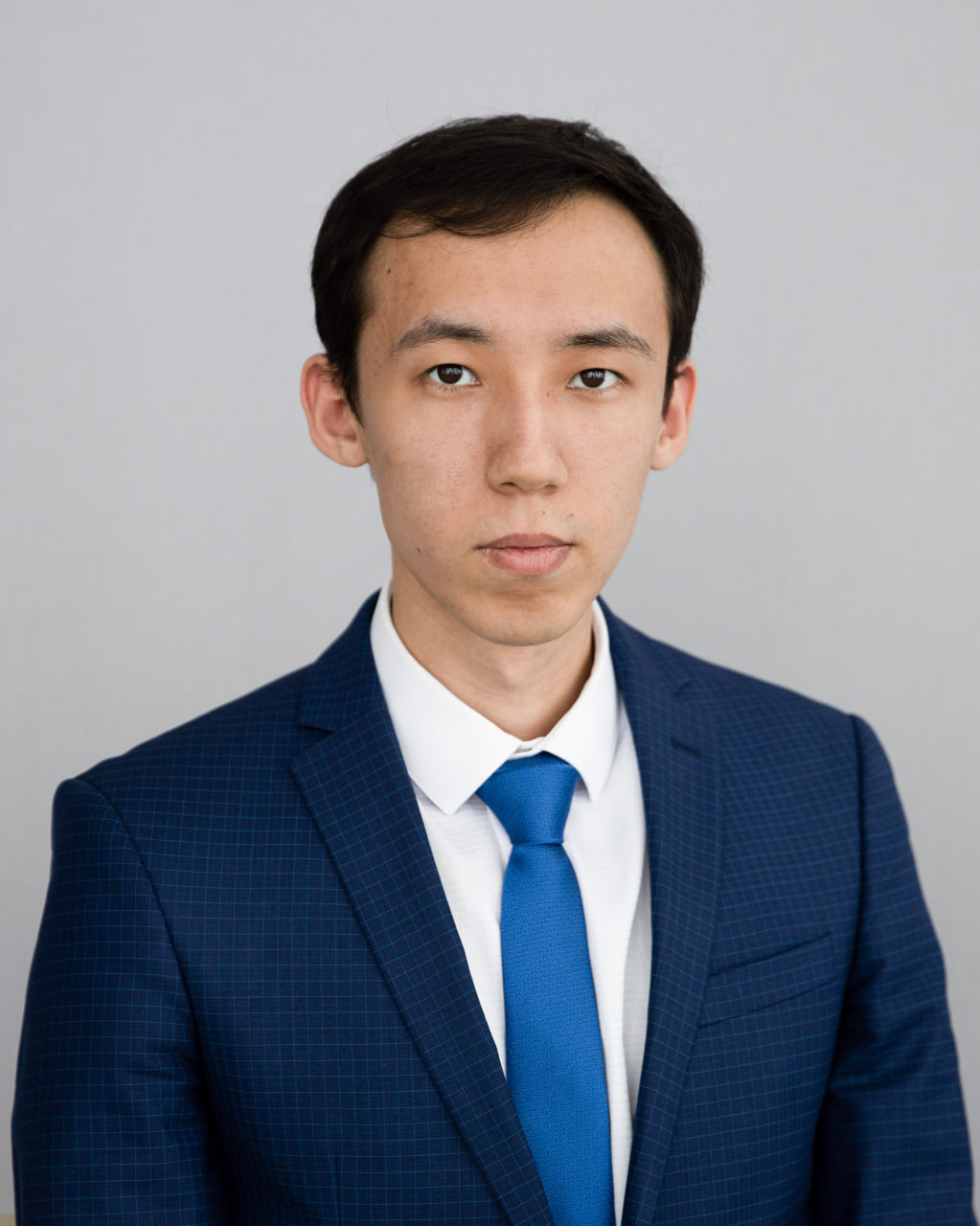}
{\bf Yernar Zhetpissov} received the BSc degree in robotics and mechatronics from Nazarbayev University, Astana, Kazakhstan, in 2020, and M.S.E. degree in robotics from Johns Hopkins University, Baltimore, MD, USA, 2023. He is currently a Ph.D student studying robotics engineering in Worcester Polytechnic Institute, Worcester, MA, USA. \\
His research interests include medical robotics and computer-integrated surgery.

\noindent\includegraphics[width=1in]{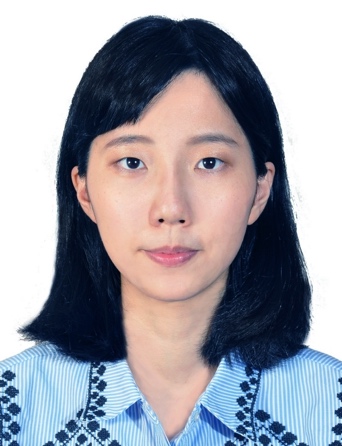}
{\bf Alexandra Cheng} 
received her B.S. degree in mechanical engineering from National Tsing Hua University and a M.S.E. in biomedical engineering from Johns Hokpins University. She is a current Ph.D student in the Johns Hopkins School of Medicine, focusing on neural implants and brain computer interfaces.

\noindent\includegraphics[width=1in]{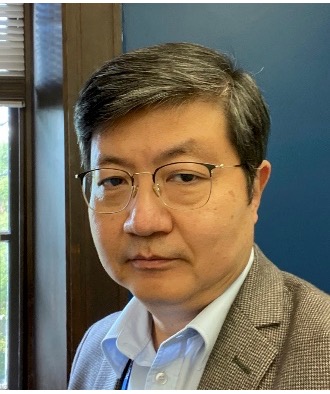}
{\bf Jin Seob Kim} 
(Member, IEEE) received the M.S. degree in Mechanical Engineering from Seoul National University, Seoul, South Korea, and the M.S and Ph.D. degrees in mechanical engineering from the Johns Hopkins University, Baltimore, Maryland, USA, in 2000, 2004, and 2006, respectively. \\
He is currently a senior lecturer in mechanical engineering and robotics with the Johns Hopkins University, Baltimore, USA. Before then, he was an assistant research professor in mechanical engineering and robotics with the Johns Hopkins University. He did postdoctoral work at the Johns Hopkins Institute for NanoBioTechnology, the Johns Hopkins Bloomberg School of Public Health, and Johns Hopkins University. His research interests include medical robotics, in particular, sensor-based mathematical modeling of continuum robots, computational modeling of biological systems, and application of non-commutative harmonic analysis.

\noindent\includegraphics[width=1in]{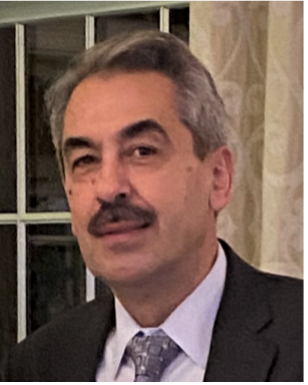}
{\bf Iulian I. Iordachita} 
(M’08-SM’14) received the M.Eng. degree in industrial robotics and the Ph.D. degree in mechanical engineering, from the University of Craiova, Craiova, Romania, in 1989 and 1996, respectively.\\
In 2000, he was a Postdoctoral Fellow with the Brady Urological Institute, School of Medicine, Johns Hopkins University, Baltimore, USA, and in 2002-2003, he was a Research Fellow with the Graduate School of Frontier Sciences, The University of Tokyo, Tokyo, Japan. He is currently a Research Professor in mechanical engineering and robotics with the Johns Hopkins University, Baltimore, USA. His research interests include medical robotics, image guided surgery with a specific focus on microsurgery, interventional MRI, smart surgical tools, and medical instrumentation.

\end{multicols}
\end{document}